\documentclass{gGAF2e}

\def\refpar{\noindent \hangindent=2em \hangafter=1}

\begin{document}

\markboth{G.A. Glatzmaier {\it et al.}}{Differential rotation in giant planets}

\title{Differential rotation in giant planets maintained by
density-stratified turbulent convection}

\author{GARY A. GLATZMAIER\thanks{$^\ast$Corresponding author. Email:
glatz@es.ucsc.edu}$^\ast\dagger$, MARTHA EVONUK$\ddagger$ and TAMARA M. ROGERS$\S$
\\ $\dagger$Earth and Planetary Sciences Department, University of California,
\\ 1156 High Street, Santa Cruz, CA 95064, USA
\\ $\ddagger$Institut f\"{u}r Geophysik, ETH Z\"{u}rich H\"{o}nggerberg,
\\ 8093 Z\"{u}rich, Switzerland
\\ $\S$High Altitude Observatory, National Center for Atmospheric Research,
\\ 3080 Center Green Drive, Boulder, CO 80301, USA}

\maketitle

\begin{abstract}
The zonal winds on the surfaces of giant planets vary with latitude.
Jupiter and Saturn, for example, have several bands of alternating eastward (prograde)
and westward (retrograde) jets relative to the angular velocity of their global
magnetic fields. These surface wind profiles are likely manifestations of the
variations in depth and latitude of angular velocity deep within the liquid interiors
of these planets.  Two decades ago it was proposed that this differential rotation
could be maintained by vortex stretching of convective fluid columns that span the
interiors of these planets from the northern hemisphere surface
to the southern hemisphere surface.
This now classic mechanism explains the differential rotation seen in laboratory
experiments and in computer simulations of, at best, weakly turbulent convection in
rotating constant-density fluid spheres.  However, these experiments and simulations
are poor approximations for the density-stratified strongly-turbulent interiors of
giant planets.  The long thin global convective columns predicted
by the classic geostrophic theory for these
planets would likely not develop.  Here we propose a much more robust mechanism for
maintaining differential rotation in radius based on the local generation of vorticity
as rising plumes expand and sinking plumes contract.  Our high-resolution
two-dimensional computer simulations demonstrate how this mechanism could maintain
either prograde or retrograde surface winds in the equatorial region of a giant planet
depending on how the density scale height varies with depth.
\end{abstract}

\noindent {\it Keywords:} Giant planet interiors, differential rotation, 
density-stratification

\section{Introduction}

The complicated flow patterns observed at the cloud tops of giant planets are likely
due to a combination of atmospheric phenomena in the shallow surface layer and thermal
convection within the deep interior.  The question, which has been debated for
several decades, is what differential rotation exists deep below the surface of
a giant planet and what dynamics maintains it.
Differential rotation is the axisymmetric zonal wind pattern in latitude and
radius relative to the deep-seated global magnetic field, which is assumed to
be rotating at approximately the mean rotation rate of the planet.  Differential rotation
can be maintained by either axisymmetric Coriolis forces arising 
from a thermally-driven meridional circulation, i.e., as a thermal wind, or by the 
convergence of Reynolds stress, i.e., the nonlinear transport of longitudinal 
momentum in latitude and radius.  Magnetic forces, like viscous forces,
typically inhibit differential rotation; however in some cases they too can
drive differential rotation (e.g. Dormy {\it et al.} 2002).

Most theoretical studies of this problem have approached it from an Earth-atmosphere
context, ignoring the dynamics of the vast interior and
using models and approximations only appropriate for shallow atmospheres.
See Dowling (1995) for a review of this approach.
These models typically assume a hydrostatic balance in
the radial direction instead of solving the full momentum equation.  That is,
they ignore the radial component of the flow in transporting
longitudinal momentum or producing Coriolis forces, two critical elements
for maintaining differential rotation in deep convective zones.
Instead, these shallow-atmosphere models of giant planets usually
rely on a thermal wind scenario (e.g. Allison 2000)
or the convergence of latitude-longitude Reynolds stress to drive zonal winds.
The latter produces a
retrograde (i.e., westward directed) zonal wind in the equatorial region
(e.g. Williams 1978, Cho and Polvani 1996) because the local
vertical component of the
planetary rotation rate (the only component these models consider) increases with
latitude.   Retrograde zonal flow in the equatorial region
is observed on surfaces of Uranus and Neptune (Hammel {\it et al.} 2005)
but a prograde (i.e., eastward
directed) equatorial jet is observed on our gas giants, Jupiter (Porco {\it et al.} 2003) and
Saturn (Sanchez-Lavega {\it et al.} 2000).  The zonal wind at higher latitudes on the gas
giants has a latitudinally banded pattern of alternating retrograde and prograde flows.
A banded pattern of zonal flows with a prograde equatorial jet can be obtained
by tuning a heating function distributed in latitude and radius
that continually nudges the temperature toward a prescribed profile that drives
the desired zonal wind pattern (e.g. Williams 2003).
However, in such a model the zonal wind profile in latitude and radius
is the result of the shallow layer assumption and the mathematical fit to
the surface observations via the ad hoc heating function;
therefore it does not provide a dynamically-consistent prediction or explanation.

It may indeed be the case that the zonal winds on giant planets are confined to
the shallow surface layers and driven by heat sources and instabilities there
without any influence from the deep convection below.  However, hydrostatic
shallow-atmosphere models assume this from the start and so are not capable
of predicting this to be the case.
The Galileo probe measured a doubling of the zonal wind speed
with depth in the surface layer of Jupiter (Atkinson {\it et al.} 1998).
A scale analysis (Ingersoll and Pollard 1982) suggests that the zonal
winds on Jupiter and Saturn extend well below their surfaces.
Probably the strongest indication of the existence of deep convection is that
spherically symmetric evolutionary models with detailed equations of state and opacities
predict convection throughout the deep fluid interiors of our gas giants
(e.g. Hubbard {\it et al.} 1999, Guillot 2005, and T. Guillot private communication).
Therefore, to predict what differential rotation exists below the surface
and to understand how it is maintained one needs a global
dynamically-consistent model that extends deep below the surface, with sufficient
physics and the full set of three-dimensional (3D) equations of motion.
The model may also need
good representations of radiative transfer and moist convection near the
surface (Ingersoll {\it et al.} 2000), a hydrogen phase transition well below
the surface separating the semi-conducting and metallic regions (Nellis 2000)
and magnetic field generation with its feedback on the flow.

An alternative to a shallow-atmosphere model is a dynamically-consistent
3D global model from the geodynamo community.  Most giant planet studies using this
approach apply the Boussinesq approximation to the set of equations that
describes deep rotating convection.  That is, a constant background density is assumed.
These 3D global models do produce banded zonal winds and a prograde equatorial
jet without neglecting the dynamics
of the deep interior and without prescribing ad hoc heating distribution functions
(e.g. Sun {\it et al.} 1993, Christensen 2002, Stanley and Bloxham 2004, Heimpel {\it et al.} 2005).
The studies find that the convergence of Reynolds stress, both latitude-longitude
and radius-longitude, dominates over the thermal wind mechanism in maintaining
differential rotation and that the
kinetic energy in the meridional circulation is typically several orders of magnitude
less than that in the differential rotation (e.g. Christensen 2002).
That is, these studies suggest that meridional circulation is
maintained by the Coriolis forces arising from the differential rotation, not the
other way around (i.e., the differential rotation is not a thermal wind).  However,
because these Boussinesq models assume a constant background density and relatively
large viscosity, the differential rotation in these simulations is maintained by the
vortex stretching of large convective columns due to the sloping impermeable
boundary.

Unlike these Boussinesq models, the interior of a giant planet has a significant
density stratification, especially near the surface, and much smaller viscosity,
which produces strong convective turbulence characterized by a
broad spectrum of scales (e.g. Glatzmaier 2005).
This turbulence is dominated by small-scale fluid parcels, which initially develop as
thermal boundary layer instabilities, then detach and either rise from the
inner boundary or sink from the outer boundary.
Their dynamical evolution in the bulk of the convection
zone is determined by nonlinear vortex-vortex interactions and the local density
stratification, not by the spherical curvature of distant boundaries.
The deep-convection constant-density models from the geodynamo community
produce flow amplitudes relatively independent of depth, opposite of that assumed
in the hydrostatic shallow-atmosphere models from the climate community.
Both, however, completely ignore the maintenance of differential rotation
by radial flow through a density stratification.

Glatzmaier and Gilman (1981, 1982), using quasi-linear solutions to
the anelastic equations of motion, examined
the role of density stratification in generating vorticity and maintaining
differential rotation by compressional torque
without requiring the vortex stretching of columnar convection.
Ingersoll and Pollard (1982) and Busse (1986) addressed the role of
density stratification using scale and asymptotic analyses.  However,
they retained Busse's original assumption of vortex-stretching by
geostrophic columnar convection spanning the entire interior,
replacing the column length by the integrated column mass.

Here we examine the deficiency of the classic vortex-stretching mechanism for
maintaining differential rotation in giant planets.
Then we describe a more robust mechanism that accounts for their large density
stratifications and strong turbulence without requiring geostrophic
columns that span the interior.  We study this mechanism with nonlinear
simulations of turbulent convection that focus on the convergence of radial-longitudinal
Reynolds stress for maintaining differential rotation in radius.

\section{The classic vortex-stretching mechanism}

3D constant-density Boussinesq models
rely on the classic vortex-stretching mechanism (Busse 1983, 2002)
to generate vorticity and maintain differential
rotation.  That mechanism requires thermal convection in the form of
vortex columns, with axes parallel to the planet's axis of rotation,
spanning from the northern to southern boundary surfaces.  Vorticity is
generated by the deformation the fluid experiences parallel to the rotation
axis as it circulates within these global columns.
The curvature of the impermeable spherical boundaries is
responsible for the longitudinal tilt of the columnar flow and the resulting
convergence of prograde angular momentum in the outer regions and retrograde
angular momentum in the inner regions.  This type of columnar convection
naturally develops when the Coriolis and pressure gradient forces nearly
balance everywhere and all other forces are relatively small, i.e., when the
flow is geostrophic.  The classic Proudman-Taylor theorem (Proudman 1916)
for incompressible fluid then predicts, to first order, that fluid
flow is within planes parallel to the equatorial plane.

However, it is unlikely that the classic vortex-stretching mechanism plays
a role in the low-viscosity turbulently-convecting interiors of giant planets
(or stars) because these columns would be extremely thin.  If they would
develop, they would quickly become unstable, buckle and shred apart into
many short unconnected vortices because nonlinear inertial effects
are not negligible at scales as small as the column diameters.  That is,
although the amplitudes of the deep convective velocities may be small,
their gradients are not.

To estimate the typical diameter of convective columns in a giant planet,
we choose Jupiter-like values for the molecular viscosity (Stevenson 1982),
$\nu = 10^{-6}$m$^2$s$^{-1}$,
planetary rotation rate, $\Omega  = 10^{-4}$s$^{-1}$, and radius,
$R = 10^8$m.  This makes the Ekman number, Ek $= \nu /(2 \Omega R^2)$, about $10^{-18}$.
If one then adopts the theoretical prediction (Roberts 1968, Busse 1970)
that the diameter of convective columns (at onset of convection)
should go like Ek$^{1/3}R$,
these columns would have diameters of roughly $100$m, a million times smaller than
their lengths.  Of course convection in giant planets is not at its onset, so
one might invoke a turbulent (eddy) viscosity many orders of
magnitude greater than the molecular value to obtain a column diameter comparable
to the latitudinal scale of the observed zonal jets.
However, even if one assumed the largest possible turbulent viscosity
equal to the product of the column diameter, $D$, and its convective velocity
normal to the axis, $v_r$,
which is estimated to be of order $10^{-3}$ms$^{-1}$ deep below the surface
based on the observed heat flux (Starchenko and Jones 2002),
the convective column diameters,

\begin{displaymath}
D \approx \left( \frac{v_r R}{\Omega} \right)^{1/2} ~,
\end{displaymath}

\noindent would be several thousand times smaller than their lengths.

Alternatively, if one acknowledges the strongly nonlinear nature of
convection in our rapidly rotating giant planets,
one would balance the generation of local non-axisymmetric vorticity
by Coriolis forces with
the advection of this vorticity.  In these constant-density models,
for which ${\bf \nabla \cdot v}$ vanishes, vorticity is assumed to be
generated by the classic vortex-stretching mechanism.
The stretching torque goes like

\begin{displaymath}
\Omega \frac{\partial v_z}{\partial z}  \approx \frac{\Omega v_r}{L_{stretch}}
\end{displaymath}

\noindent where $z$ and $r$ are the cylindrical
coordinates parallel and perpendicular to ${\bf \Omega}$, respectively, and
$v_z$ is the secondary fluid flow parallel to the column (i.e., the stretching velocity).
$L_{stretch}$ is the geometric length scale in the $r$ direction
for column stretching due to the spherical boundaries; it is equal to
$|d~ \mathrm{ln} H/dr|^{-1} = (R^2-r^2)/r$,
where $H$ is the column length, $2(R^2-r^2)^{1/2}$ (Busse, 2002).
$L_{stretch}$ varies considerably with latitude;
it is approximately the planetary radius for columns intersecting the
surface at mid-latitude.
This vorticity generation would be balanced with the advection by velocity
${\bf V}$ of vorticity ${\bf \nabla \times v}$, which
has an approximate amplitude of $V v_r / D^2$.
The balance gives the Rhines-like scale (Rhines 1975, Heimpel {\it et al.} 2005)
for the column diameters of

\begin{displaymath}
D \approx \left( \frac{V L_{stretch}}{\Omega} \right)^{1/2} ~~~~~~~~ {\rm and}
~~~~~~~~ \frac{D}{H} \approx \left( \frac{V}{4 \Omega r} \right)^{1/2}.
\end{displaymath}

\noindent This assumes the banded zonal wind, i.e. the differential rotation
which acts to destroy the columnar flow structures by shearing them,
is balanced by their generation via vortex stretching.
If one chooses $V$ to be the maximum zonal wind speed on Jupiter's surface,
$100$ms$^{-1}$, the diameter of columns intersecting the surface at low latitude
would be about a hundred times smaller than $R$ and about twenty times smaller than
their lengths.

These vortex-stretching arguments have been based on the constant-density
approximation for a giant planet's interior and an impermeable outer boundary.
However, unlike laboratory experiments and previous 3D computer models,
the surfaces of giant planets are not strictly impermeable.
It is unlikely that flows in the low-density highly-compressible
strongly-turbulent surface region of a giant planet could drive the
$z$-directed secondary flow in the much denser fluid deep below the surface
required for the vortex-stretching mechanism.

If one does acknowledge the large density stratifications
in our giant planets, local vorticity is easily generated by compressional torque:

\begin{displaymath}
- 2 {\bf \Omega  \nabla \cdot v} ~~.
\end{displaymath}

\noindent This is proportional to the
radial component of convective velocity ($v_r$) and inversely proportional to the
local density scale height, $L_{compress}$ (see Appendix A).
If, at low latitude, the amplitude of vorticity were again
approximated as convective velocity ($v_r$) divided by the vortex diameter,
balancing this compressional torque with the advection of vorticity
would make the typical vortex diameter be

\begin{displaymath}
D \approx \left( \frac{V L_{compress}}{\Omega} \right)^{1/2} .
\end{displaymath}

\noindent Since the local density scale height is very small
in the outer region of Jupiter
(Bodenheimer {\it et al.} 2003, Guillot 2005, T. Guillot private communication)
the characteristic diameter of vortices in the shallow equatorial region
would be a thousand times smaller than Jupiter's radius $R$.
However, unlike the geostrophic vortex-stretching mechanism,
this compressional-torque mechanism
does not require vortices spanning from
the northern to the southern boundaries.  Small vortices aligned with
the planet's rotation axis likely {\it do} exist within its turbulent interior;
but most are likely short isolated structures.

One might imagine geostrophic columns close to the planetary
rotation axis.  Fluid circulating in such columns at low latitude (i.e.,
in the deep interior) would span only a fraction of a density scale height.
At high latitude, where the density scale height is small, the flow
would be mainly in latitude; therefore again fluid parcels would experience
little density stratification.  However, the relative change in volume
experienced by fluid parcels circulating in cylindrical radius
within such columns would vary along the length of the column,
depending on the parcel's latitude and on the planet's density profile
in spherical radius.  This will generate vorticity via compressional torque
at rates that vary along the column length and therefore would likely cause the
columns to buckle and shear apart.

In this simple scale analysis we have been assuming the cross sections of
vortices are roughly circular.  However, a significant differential rotation
would likely shear them into vortex sheets.
In such a scenario their longitudinal dimensions and velocities would be
much larger than those in cylindrical radius.

\section{The density-stratification mechanism}

The alternative mechanism proposed here for maintaining differential rotation in
giant planets (and stars) is based on compressional torque instead of
vortex stretching torque.  Consider a buoyant fluid parcel rising through
a density-stratified medium.  As it expands it produces a Coriolis torque that causes
it to rotate in the opposite direction of the basic rotation of the planet.  
Note that the Coriolis force due to just the rise of the parcel normal to
the rotation axis is balanced by part
of the pressure gradient force and so does not generate vorticity;
it is the Coriolis torque due to the expansion
of the parcel as it rises (i.e., the divergence of the velocity parallel to
the equatorial plane) that generates negative vorticity (i.e., an anticyclone)
relative to the rotating frame of reference (see equation A4).
Likewise, the contraction of a sinking fluid parcel generates positive
vorticity (i.e., a cyclone).

One can also describe this process by defining
the potential vorticity of the fluid parcel
as the total vorticity (twice the planet's rotation rate plus its local vorticity
within the rotating frame) divided by its density and can show that this quantity
is approximately conserved as the parcel moves
(Ertel 1972, Glatzmaier and Gilman 1981).  That
is, as the density of a rising fluid parcel decreases, its local vorticity also
decreases, and vice versa.  The derivation of this theorem (see Appendix B) includes
the nonlinear inertial term in addition to the pressure gradient and Coriolis terms
in the momentum equation and therefore is more appropriate for describing rotating
turbulent flows than the linear steady-state Proudman-Taylor theorem.

Turbulent convection clearly does not have the traditional
large-scale cellular structure seen in laminar convection.
Instead, one needs to consider the motion and deformation of small isolated plumes
that gain and lose vorticity as they sink and rise, respectively.
The density-stratification mechanism for maintaining differential rotation is
based on the resulting tilted flow trajectories which transport angular momentum from
one region to another.

\subsection{Two models with different radial profiles of density}

As a simple demonstration, we present two numerical simulations of strongly turbulent
thermal convection in the equatorial plane of a rotating density-stratified planet.
The model is two-dimensional (2D), i.e., there are no flows or gradients
perpendicular to the plane of the rotating disk.  It allows convection down to
a solid core at 20\% of the outer boundary.   To keep this model simple, the
outer boundary is set at the radius where the density is just seven times smaller
than that at the inner boundary.
The anelastic equations of motion are solved using a
spectral semi-implicit method (see Appendix A).  The spatial resolution is 4096
Fourier grid points in longitude by 1537 Chebyshev grid levels in radius.
Each case was run for more than 2,000,000 numerical time steps, representing
more than 1000 rotation periods.

\begin{figure}
\centerline{\epsfbox{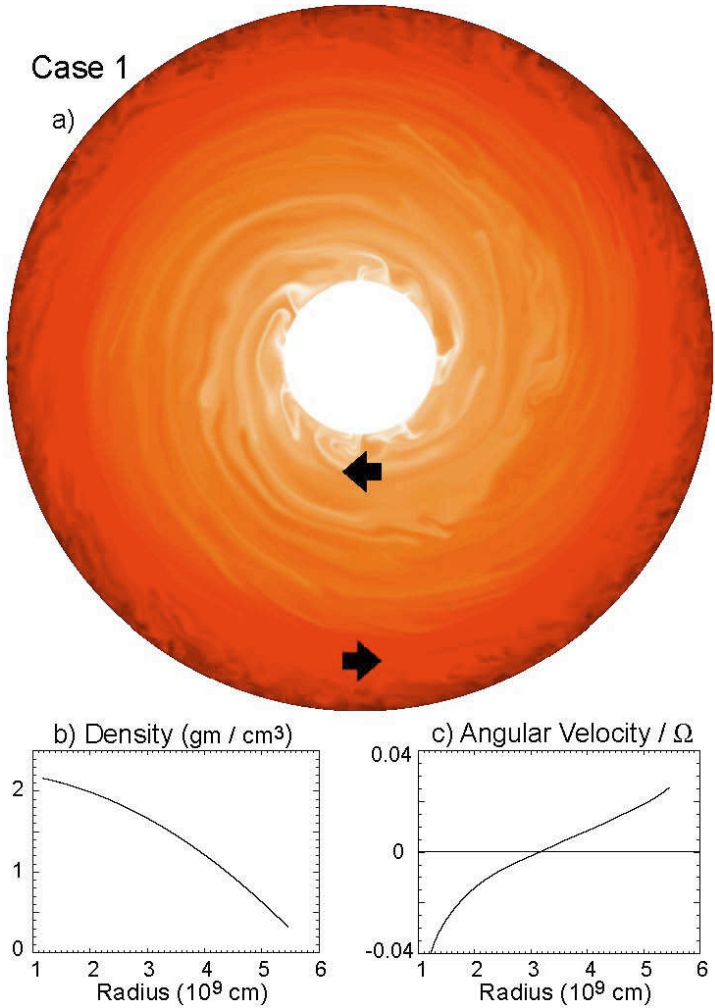}}
\caption{2D convection and differential rotation, case 1.
a. A snapshot of the thermal (entropy)
perturbations in a 2D model of the equatorial plane of a giant planet.  This is
viewed in the rotating frame, down from the northern hemisphere; the planetary
rotation is counter-clockwise.  The arrows indicate the direction of the zonal
flow relative to the rotating frame; the clockwise-directed arrow represents
negative (retrograde) angular velocity.  Light colors represent hot, buoyant fluid;
dark colors represent cold, heavy fluid.  b. The prescribed reference state density
profile.  c. The time averaged angular velocity relative to the rotating frame
and scaled by the planetary rotation rate.}
\end{figure}

\begin{figure}
\centerline{\epsfbox{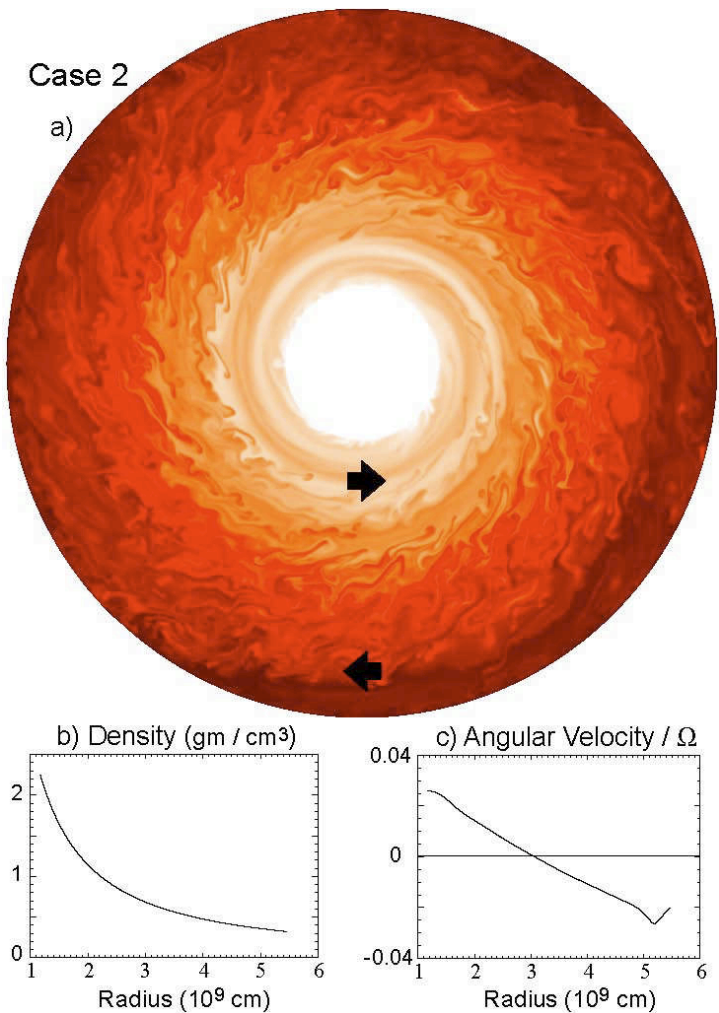}}
\caption{2D convection and differential rotation, case 2.  As in figure 1.}
\end{figure}

Several traditional nondimensional parameters characterize these simulations:

\begin{displaymath}
\mathrm{Ra} = \frac{g |\partial \rho / \partial S | \Delta S D^3}
{\rho \nu \kappa} ~~~,~~~~~~~~~
\mathrm{Ek} = \frac{\nu}{2 \Omega D^2} ~~~,~~~~~~~~~
\mathrm{Pr} = \frac{\nu}{\kappa} ~~~.
\end{displaymath}

\noindent Here, $g$ is gravitational acceleration, $\rho$ is density,
$S$ is specific entropy, $\Delta S$ is the drop in $S$ across the depth $D$ of
the convection zone, $\nu$ is viscous diffusivity, $\kappa$ is thermal
diffusivity and $\Omega$ is the planetary rotation rate.
Values of these variables at mid-depth are used in the nondimensional numbers.
The Rayleigh number (Ra), a measure of convective driving
relative to viscous and thermal diffusion, is $3 \times 10^{11}$ and the
Ekman number (Ek), a measure of viscous to Coriolis effects, is $10^{-8}$.
The Prandtl number (Pr), viscous relative to thermal diffusion, is $0.2$.
A convenient combination of these numbers, the convective Rossby number (Rc),
estimates buoyancy to Coriolis effects without considering diffusive effects:

\begin{displaymath}
\mathrm{Rc = \left( \frac{Ra}{Pr} \right)^{1/2} Ek} ~~~.
\end{displaymath}

\noindent In these simulations Rc is about $10^{-2}$;
i.e., Coriolis forces are typically
a hundred times greater than buoyancy forces.

The resulting zonal wind amplitude, $V$,
relative to the rotating frame of reference is a few hundred thousand
times greater than the viscous diffusion velocity (i.e.,
the Reynolds number (Re) is roughly $10^5$) and is a hundred times smaller than
the velocity of the rotating frame (i.e., the Rossby number (Ro) is $10^{-2}$),
where

\begin{displaymath}
\mathrm{Re} = \frac{V D}{\nu} ~~~,~~~~~~~~~
\mathrm{Ro} = \frac{V}{2 \Omega D} ~~~.
\end{displaymath}

\noindent Convective velocities are typically two orders of magnitude smaller than the
surface zonal winds they maintain.

These test cases are ideal because if there were no density stratification in this 2D
configuration the Coriolis forces would be completely balanced by part of the
pressure gradient (see equation A4)
and would therefore generate no vorticity and no differential rotation.
Also, since there is no flow perpendicular to the 2D plane the longitudinally
averaged radial flow vanishes and so there is no thermal wind effect.
Therefore, the differential rotation maintained in these simulations is driven entirely by
the density-stratification mechanism.  The two cases are given the same control
parameters and the same total density drop (a factor of seven) across the convection
zone but different radial profiles of density.  The relative change in density
with radius is greatest in Case 1 (figure 1b) near the outer boundary;
whereas in Case 2 (figure 2b) it is greatest near the inner boundary.

The density-stratification mechanism for
maintaining differential rotation involves a radially-dependent phase
propagation in longitude of the vorticity pattern, which tilts the
plume trajectories in longitude and thus redistributes angular
momentum in radius.  To undersand how this happens think of the series
of upflows and downflows at the onset of convection as convection cells.
Positive vorticity peaks in the centers of counter-clockwise circulating
cells (cyclones) and negative vorticity peaks in the centers of
clockwise circulating cells (anticyclones), when viewed in the rotating frame
from the north.  However, the rate that vorticity is being generated peaks
in the upflows and downflows between the cell centers, i.e., $90^\circ$ out
of phase relative to the existing pattern of vorticity.

Since rising fluid generates anticyclonic vorticity
on the prograde side of anticyclones and on the retrograde side
of cyclones, the phase of this circulation pattern propagates in the
prograde direction (Glatzmaier and Gilman 1981).  This Rossby-like vorticity
wave occurs because density decreases with radius.  Consider Case 1.
Since the relative decrease in density with radius {\it increases} with
radius in this case, the pattern of circulation propagates eastward
faster at greater radii.  This causes rising fluid to tilt eastward (i.e.,
prograde) and sinking fluid to tilt westward (i.e., retrograde).
That is, prograde angular momentum is
transported toward the outer boundary and retrograde angular momentum is
transported away from the outer boundary.  The opposite occurs at the
inner boundary.  This convergence of Reynolds stress maintains
prograde zonal flow in the outer part of the convection zone and
retrograde flow in the inner part (figure 1c).
Once the differential rotation is established
it provides positive nonlinear feedback.   That is, rising and sinking plumes
are sheared by the differential rotation, which increases the tilt and
tightens the spiral flow structure. 
For Case 2, the relative decrease in density with radius
{\it decreases} with radius; therefore the eastward phase propagation is faster
at greater depth, which establishes the opposite spiral tilt and differential
rotation (figure 2).

Notice that the flow in Case 1 (figure 1a) appears much less turbulent than that in
Case 2 (figure 2a).  This is likely related to the {\it Rayleigh Stability Criterion}
for a differentially-rotating inviscid fluid disk.  For centrifugal
force not to exceed gravitational and pressure gradient forces when a fluid parcel
is displaced outward, the radial gradient of specific angular momentum in the disk,
$r^2 \Omega (r)$, needs to be positive.  Here, $\Omega(r)$ is the total
angular velocity of the fluid disk in the inertial frame.
Although this stability criterion is satisfied everywhere for both cases, the positive
gradient of $\Omega(r)$ for Case 1 (figure 1c) makes it more stable
than Case 2 (figure 2c); that is, Case 1 provides greater restoring forces
against radial displacements.

The evolution of turbulent buoyant parcels detaching from the boundary layers, being
differentially torqued by Coriolis forces due the their expansion or contraction, and
ultimately establishing a strong differential rotation is clearly seen in computer
graphical movies of the two cases presented here
(http://es.ucsc.edu/~glatz/diskmovies).  The structure of this highly-turbulent
rotationally-dominated flow is a superposition of small-amplitude non-axisymmetric
radial motions (figures 1a and 2a) and the large-amplitude axisymmetric differential
rotation (figures 1c and 2c); the result is a series of wave-like flows in longitude.
When viewed in the rotating frame of reference, defined as the one with zero
total angular momentum, the differential rotation continually winds the thermal
spiral patterns.  On average, the inner and outer regions make one revolution in
opposite directions about every fifty planetary rotation periods.  This rate likely
depends on the convective Rossby number, Rc.

The robust feature of the density-stratification mechanism presented here is that it
acts locally.  It generates vorticity and maintains differential rotation even in a
strongly turbulent and magnetic environment without requiring
vortex columns spanning the entire interior.

Unlike Jupiter and Saturn, the ice giants, Uranus and Neptune, have westward
(retrograde) equatorial zonal winds at their cloud tops, which may be maintained by
shallow atmospheric flows (e.g. Williams 1978, Cho and Polvani 1996).
Alternatively, a density profile with a steep
gradient well below the surface at the liquid-gas transition, as predicted in models
of Uranus and Neptune (Hubbard {\it et al.} 1991), might
maintain the westward equatorial zonal wind at the surface as in figure 2
if the density scale height at the transition were smaller than that near the
surface.

\subsection{A fully convective model with no solid core}

One-dimensional evolutionary models of Jupiter cannot definitely say that
this planet has a rocky core (Hubbard {\it et al.} 1999, Guillot 1999).
It is also likely that some extra-solar giant planets, at least during parts
of their lifetimes, are fully convective with no solid core
(Bodenheimer {\it et al.} 2003, Guillot 2005).
Therefore it is interesting to compare the flow structures predicted
for fully convective giant planets to those with rocky cores.

\begin{figure}
\centerline{\epsfbox{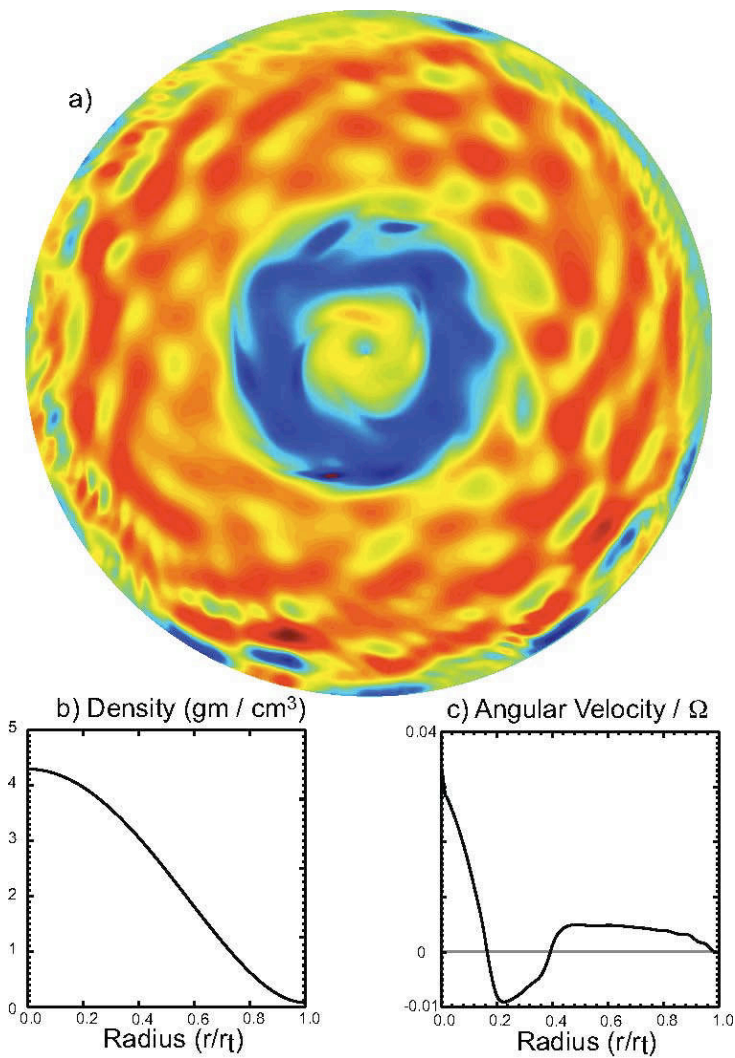}}
\caption{2D convection and differential rotation without an inner core.
a. A snapshot of the
angular velocity in a 2D model of the equatorial plane of a giant planet, viewed in
the rotating frame with planetary rotation counter-clockwise.
Red and yellow indicate prograde angular velocity; blue is retrograde.
This case shows three alternating bands of zonal flow.
b. The prescribed reference state density profile.
Radius is indicated as a fraction of the outer boundary radius.
Density at the center is 54.6 times greater than that
at the outer boundary.
c. The time averaged angular velocity relative to the rotating frame
and scaled by the planetary rotation rate.}
\end{figure}

Numerical simulations using a 2D anelastic finite-volume model of the
equatorial plane of a giant planet with no solid core also demonstrate
that differential rotation in radius is maintained by the density-stratification
mechanism (Evonuk and Glatzmaier 2006, 2007).
These studies illustrate how the structure of the convection and zonal flow are
significantly influenced by the presence or absence of a solid core when the
effects of planetary rotation are small (i.e., for large convective Rossby number, Rc).
In particular, a dominant dipolar flow is preferred; that is, a flow straight
through the center 
since this is the most efficient way of removing heat from the central region.
However, when Coriolis forces are relatively significant (small Rc)
the flow is dominated by the zonal flow.

Simulations with this model (Evonuk 2008) have also shown that the
number of alternating bands (in radius) of angular velocity
(relative to the mean rotation rate) increases as the
Rc decreases.  Figure 3 illustrates a case with
three alternating bands in radius.  The parameters for
this case (with $D$ being the full radius) are
Ra = $10^{12}$, Ek = $10^{-7}$, Pr = 1 and therefore Rc = 0.1.
The reference state density at the center is 54.6 times greater than that
at the outer boundary (figure 3b); its radial derivative vanishes at both the center and
at the outer boundary.  The density scale height is minimum
in the outer part of the convection zone, which maintains prograde
zonal flow there and retrograde flow below (figures 3a,c).
The central region has significant prograde angular velocity
because the advection of vorticity through this region is by
the dipolar flow (i.e., longitudinal wavenumber one).  This brings in
prograde vorticity (since sinking fluid generates positive vorticity) and
removes retrograde vorticity (since rising fluid generates negative vorticity).
Similar simulations but with slightly larger Rc produce only two alternating
bands; simulations with much larger Rc produce a dipolar flow structure.

\subsection{A model with a stable interior below a convection zone}

\begin{figure}
\centerline{\epsfbox{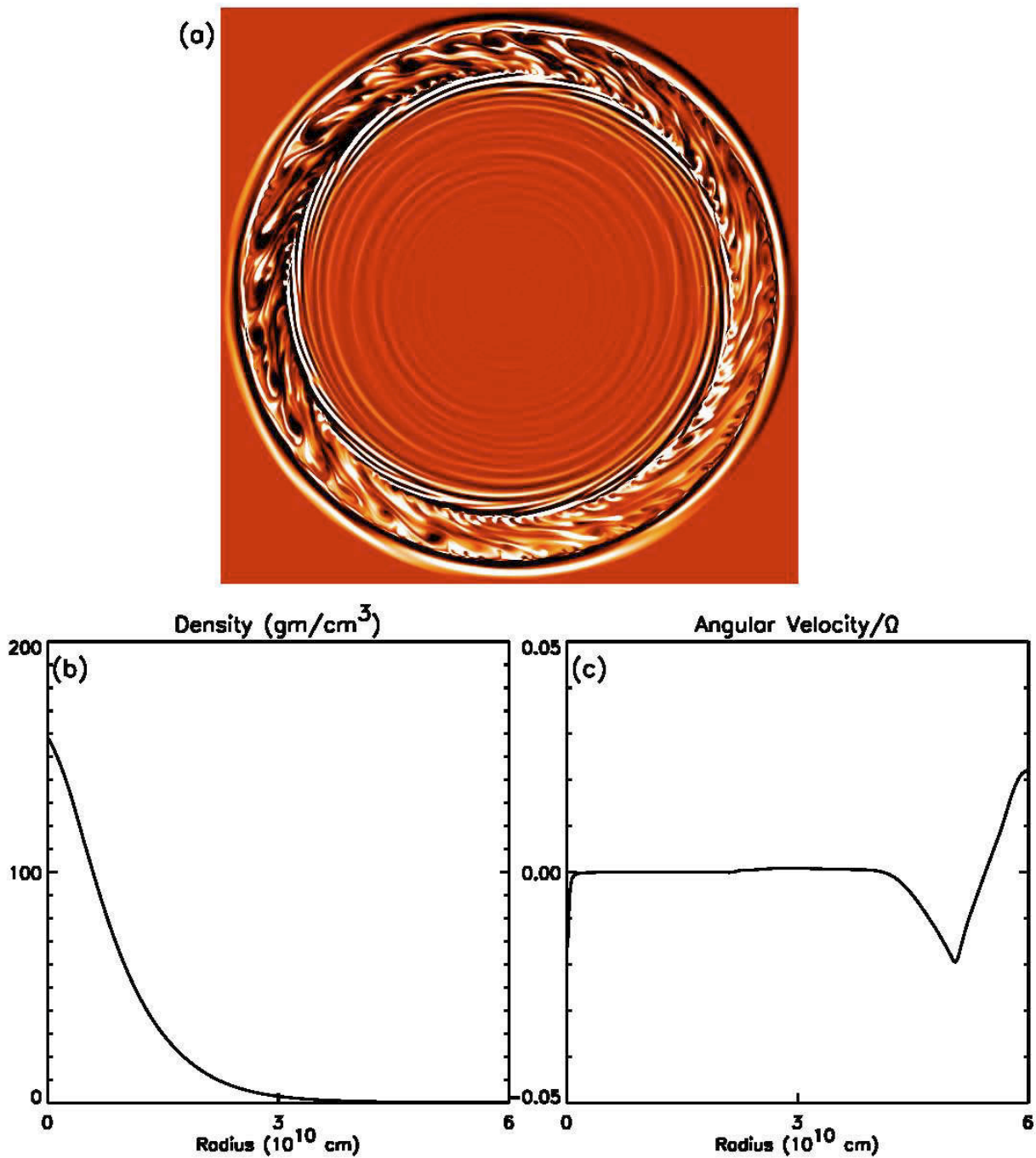}}
\caption{a. A snapshot of the temperature perturbations (relative to the
radially-dependent reference state temperature)
in a 2D model of rotating convection in the equatorial plane.  As in the
other figures, rotation is counter-clockwise.
Convection exists in the outer unstable (superadiabatic) region
and gravity waves in the inner stable (subadiabatic) region.
Light colors represent hot, buoyant fluid;
dark colors represent cold, heavy fluid.  b. The prescribed reference state density
profile; density varies by a factor of 20 within the convection zone.
c. The time averaged angular
velocity relative to the rotating frame and scaled by the planetary rotation rate.}
\end{figure}

It is likely that some extra-solar giant planets
have a convectively stable interior below an
unstable convection zone, as does the sun.
To study such a scenario, we use another 2D anelastic model within the
equatorial plane (Rogers {\it et al.} 2006).
This model employs a spectral method in longitude and
a finite-difference method in radius.  Only the outer 20\% in radius is
convectively unstable; a very stable (subadiabatic) radiative region is
prescribed below the convection zone. 
A shallow stable layer is also prescribed above the unstable convection zone.
The reference state density, pressure and temperature are fitted
to a one-dimensional solar evolutionary model.  The radii of the model's
inner boundary, its stable-unstable interface and its outer boundary
are 0.05$R_{sun}$, 0.71$R_{sun}$ and 0.85$R_{sun}$, respectively,
where $R_{sun}$ is the radius of the solar photosphere.  The corresponding
densities at these radii are 153, 0.20 and 0.01 gm cm$^{-3}$.
The Ra, Ek and Pr depend on radius; within the lower part of the convection
zone these are $10^8$, $10^{-6}$ and $10^{-2}$, respectively.

A snapshot of the temperature perturbation in the equatorial
plane from one of these simulations is illustrated in figure 4,
showing tilted plumes in the convection zone and gravity waves in
radiative interior.  As in the previous cases, differential rotation in the
convection zone is maintained by the density-stratification mechanism.
Since the density scale height is smallest near the
surface in this model as it is in the sun, eastward (prograde) zonal flow is
maintained near the surface and westward (retrograde) flow near the base
of the convection zone.

Kinetic energy and angular momentum are transferred from downwelling
plumes in the convection zone to gravity waves in the stable interior.
The dissipation of these waves contributes to the maintenance of a
zonal flow in the upper part of the stable interior (figure 4c).
This transfer of energy and momentum to the interior reduces the
amplitude of the zonal flow in lower part of the convection zone
compared to what would be maintained by a stress-free impermeable boundary
at the base of the convection zone (Rogers and Glatzmaier 2005).

\section{Discussion}

It is interesting to compare and contrast the classic vortex-stretching
mechanism (Busse 1983, 2002) with the density-stratification mechanism described here.
Vorticity is generated in both by Coriolis torques that occur as fluid diverges
in planes parallel to the equatorial plane.
This occurs for the vortex-stretching scenario because of conservation
of fluid volume; that is, fluid in geostrophic convective columns
(spanning from the impermeable
outer boundary in the northern hemisphere to that in the southern hemisphere)
is forced to spread out as it circulates away from the axis of rotation due to
sloping ends of the columns determined by
the curvature of the spherical boundary.  For the density-stratification
mechanism, on the other hand, a compressible fluid parcel expands as it rises
because of the decrease in the local pressure.

In addition, both mechanisms require the flow trajectories to be tilted in
longitude to achieve a convergence of prograde angular momentum in one
region and a convergence of retrograde angular momentum in another.
For the vortex-stretching mechanism, prograde zonal flow is maintained
where the natural logarithm of
{\it column length} decreases most rapidly with {\it cylindrical} radius.
For a spherical outer boundary (outside the tangent cylinder to the inner
core) this occurs in the outer part of a convective column, which causes
upflow to tilt in the prograde direction and downflow to tilt in
the retrograde direction.  In particular, this mechanism always produces
a prograde zonal flow near the spherical surface in the equatorial region.
Note that a retrograde equatorial surface flow could be maintained in a rotating
{\it cylinder} filled with a constant density fluid if the ends have a
{\it concave} shape because eastward-propagating vorticity waves
would travel faster near the inner boundary, causing convective columns
to be tilted in the opposite direction (Busse 2002).
For the density-stratification mechanism, prograde flow is
maintained within a convecting region where the natural logarithm of {\it density}
decreases most rapidly with {\it spherical} radius, which is typically in the
outer part of a stratified convecting region for a gas giant.
Note, however, that the retrograde tilted upflow
and retrograde equatorial zonal flow of our Case 2 (figure 2), which
we produce with a ``concave-shaped" density profile, is similar in principle
to Busse's (2002) example of ``concave-shaped" cylinder boundaries.
Therefore, one of the differences between the two mechanisms is that
the resulting differential rotation depends on the geometry of the inner and outer
impermeable boundaries for the vortex-stretching mechanism;
whereas, for the density-stratification mechanism the resulting differential
rotation depends on the internal density structure of the planet.

This difference between the two mechanisms is particularly
critical for strongly turbulent convection, like that in a giant planet.
The vortex-stretching mechanism assumes nonlinear Reynolds stresses are too
small to produce instabilities and therefore assumes
straight convective columns with sloping
(convex) ends in contact with the impermeable
outer boundaries.  However,
in a strongly turbulent environment these long thin columns would
likely not remain straight and intact if they ever do develop.
Instead many short disconnected vortices or vortex sheets would likely exist,
most not in contact with the outer boundary.  Without the organized vortex
stretching of the classic mechanism, a stable differential
rotation profile may be difficult to maintain in a turbulent
constant-density fluid.  However, short isolated vortices
in a turbulent density-stratified environment can easily maintain a stable
differential rotation profile because this mechanism acts locally.

The degradation of differential rotation as convection, in a fluid with a
constant background density, becomes increasingly turbulent has been seen in laboratory
experiments (e.g. Hart {\it et al.} 1986, Aubert {\it et al.} 2001) and
in 3D numerical simulations (e.g. Christensen 2002).  As the convective Rossby
number, Rc, increases convection becomes more turbulent and the fluid velocities
that compose the Reynolds stress become less correlated, reducing the
kinetic energy of the differential rotation relative to that of the convection.
Christensen (2002) finds that
a fairly stable differential rotation profile can however be maintained even when
the convective columns and the resulting Reynolds stresses are intermittent in
space and time, if there is a robust time-averaged vortex-stretching mechanism.
The de-correlation in the velocities seen in these
Boussinesq simulations occurs less near the surface where
vortices still ``feel" the sloping impermeable boundary.
As mentioned, vortices in a turbulent density-stratified environment do not need to
be deformed by the boundary to maintain a differential rotation.

Several differences between Boussinesq and density-stratified convection
also exist in the details of the flow structures.
It is easy to see that the global circulation of fluid parcels that expand by
several orders of magnitude while rising would have a significantly different
pattern than the circulation of constant density fluid.
For example, consider the difference between the cylindrical geometry of
the vortex-stretching mechanism and the spherical geometry of the
density-stratification mechanism.
A secondary flow toward the equatorial plane forced by a constant-density
convecting column between sloping boundaries causes the fluid to spread out
and therefore produces negative vorticity.
However, a similar flow trajectory in a density-stratified environment
causes a fluid parcel to contract (because spherical radius decreases)
and therefore produces positive vorticity.
In the polar regions vorticity generation due to a secondary flow
toward the equatorial plane
is vanishingly small for the Boussinesq vortex-stretching scenario.
However, for the density-stratified scenario fluid rising and sinking parallel
to the axis of rotation in the polar region expands and contracts as much as it does
when flowing within the equatorial plane (over the same spherical radii)
and therefore produces considerable vorticity.
These differences in flow patterns are particularly important at shallow depths
below the surface of a giant gas planet where the volume of a
fluid parcel changes significantly as the parcel flows through many
density scale heights.

Some high-resolution 3D Boussinesq simulations
of convection and differential rotation in giant gas planets (e.g. Heimpel {\it et al.} 2005)
produce surface zonal winds that appear amazingly realistic.
However, because of their neglect of density stratification,
the structure of
convection and zonal winds below their model's surface may not be as realistic.
A convenient test (Glatzmaier 2008)
would be to specify Jupiter's radius and basic rotation period
as length and time scales
and produce surface zonal winds with roughly the same amplitude and
pattern observed on Jupiter.  Then, using a realistic depth-dependent
electrical conductivity for this semi-conducting region (Liu {\it et al.} 2008)
and the model's subsurface fluid flow
generate a magnetic field similar in amplitude and structure (above the surface)
to Jupiter's observed field.  A necessary condition for the subsurface flow
to be realistic is to be able accomplish this without generating more Ohmic
heating than Jupiter's observed luminosity.

\section{Conclusions}

We have argued that the extremely thin convective columns required to
span from the northern to southern boundaries in the classical
vortex-stretching mechanism would likely not develop in the turbulent
interiors of giant planets.  As an alternative, we have investigated
the maintenance of differential rotation due to
the effects of rising fluid expanding and sinking fluid contracting
within a density-stratified equatorial plane of a giant planet.
Our simple 2D computer simulations illustrate
how this density-stratification mechanism can maintain differential rotation
by compressional torques acting locally on small convective plumes.
We have discussed differences in the flow structures produced with constant-density
(Boussinesq) models and those produced by density-stratified (anelastic) models,
which are most significant in the outer region of a giant planet where the
density scale heights are smallest.

The opposite differential rotation profiles
illustrated in figures 1 and 2 suggest that the oppositely directed equatorial
winds observed on our ice giants compared to those on our gas giants
might be due to their different radial profiles of density
(Hubbard {\it et al.} 1991, Guillot 2005).  Our models also demonstrate how
this mechanism can maintain differential rotation in radius when the
interior is fully convective (i.e., no solid core)
and within a convection zone above a convectively stable interior.

The 2D models we have presented here, however, are meant to simply demonstrate
this fundamental mechanism, which has been neglected in most previous models
of giant planets; they are not
meant to predict the pattern or extent of differential rotation below the surface
of a giant planet.  Our Cases 1 and 2 span only two density scale heights, not
the much larger number which exist within a giant planet,
most of which occur in the shallow
layers where the density-stratification mechanism is most effective.
We have also neglected the hydrogen phase
transition, which is predicted to exist at roughly 90\% of Jupiter's radius
and 50\% of Saturn's radius (Nellis 2000), and the magnetic field that is generated
in the outer semi-conducting region where the electrical conductivity rapidly
increases with depth.  In addition, we have not addressed the maintenance of
differential rotation in latitude, which would require a 3D global simulation with
a density stratification in spherical radius.

\bigskip
\noindent{\bf Acknowledgments}
\smallskip

We thank F. Busse, U. Christensen, C. Jones, P. Olson, P. Roberts and D. Stevenson
for discussions.  T.R. is supported by an NSF Astronomy and Astrophysics
Postdoctoral Fellowship under award 0602023.
Support for this research was provided by grants from the
NASA {\it Planetary Atmospheres Program} (NAG5-11220),
the NASA {\it Outer Planets Research Program} (NNG05GG69G),
the NASA {\it Solar and Heliospheric Physics Program} (NNG06GD44G)
and from the {\it Institute of Geophysics and Planetary Physics}
at Los Alamos National Laboratory and the University of California Santa Cruz.
Computing resources were provided by NSF at the
{\it Pittsburgh Supercomputing Center} and by an MRI funded Beowulf cluster at
UCSC (AST-0521566), by the NASA {\it Advanced Supercomputing Division}
and by DOE at the {\it National Energy Research Scientific Computing Center}.

\bigskip
\noindent{\bf References}
\smallskip

\refpar
Allison, M.,
A similarity model for the windy jovian thermocline.
{\it Planet. Space Sci.}, 2000, {\bf 48}, 753-774.

\refpar
Atkinson, D.H., Pollack, J.B. and Seiff, A.,
The Galileo probe doppler wind experiment:
Measurement of the deep zonal winds on Jupiter.
{\it J. Geophys. Res.}, 1998, {\bf 103}, 22911-22928.

\refpar
Aubert, J., Brito, D., Nataf, H.-C., Cardin, P. and Masson, J.-P.,
A systematic experimental study of rapidly rotating spherical convection
in water and liquid gallium.
{\it Phys. Earth Planet. Inter.}, 2001, {\bf 128}, 51-74.

\refpar
Bodenheimer, P., Laughlin, G. and Lin, D.N.C.,
On the radii of extrasolar giant planets.
{\it Astrophys. J.}, 2003, {\bf 592}, 555-563.

\refpar
Braginsky, S.I. and Roberts, P.H.,
Equations governing core convection and the geodynamo.
{\it Geophys. Astrophys. Fluid Dyn.}, 1995, {\bf 79}, 1-97.

\refpar
Busse, F.H.,
Thermal instabilities in rapidly rotating systems.
{\it J. Fluid Mech.}, 1970, {\bf 44}, 441-460.

\refpar
Busse, F.H.,
A model of mean zonal flows in the major planets.
{\it Geophys. Astrophys. Fluid Dyn.}, 1983, {\bf 23}, 153-174.

\refpar
Busse, F.H.,
Asymptotic theory of convection in a rotating, cylindrical annulus.
{\it J. Fluid Mech.}, 1986, {\bf 173}, 545-556.

\refpar
Busse, F.H.,
Convective flows in rapidly rotating spheres and their dynamo action.
{\it Phys. Fluids}, 2002, {\bf 14}, 1301-1314.

\refpar
Cho, J.Y.-K. and Polvani, L.M.,
The morphogenesis of bands and zonal winds in the
atmospheres on the giant outer planets.
{\it Science}, 1996, {\bf 273}, 335-337.

\refpar
Christensen, U.R.,
Zonal flow driven by strongly supercritical convection
in rotating spherical shells.
{\it J. Fluid Mech.}, 2002, {\bf 470}, 115-133.

\refpar
Dormy, E., Jault, D. and Soward, A.M.,
A super-rotating shear layer in magnetohydrodynamic spherical Couette flow.
{\it J. Fluid Mech.}, 2002, {\bf 452}, 263-291.

\refpar
Dowling, T.E.,
Dynamics of jovian atmospheres.
{\it Ann. Rev. Fluid Mech.}, 1995, {\bf 27}, 293-334.

\refpar
Ertel, H.,
Ein neuer hydrodynamischer Wirbelsatz.
{\it Meteorolol. Z.}, 1972, {\bf 59}, 277-281.

\refpar
Evonuk, M.,
The role of density stratification in generating zonal flow
structures in a rotating fluid.
{\it Astrophys. J.}, 2008, {\bf 673}, 1154-1159.

\refpar
Evonuk, M. and Glatzmaier, G.A.,
A 2D study of the effects of the size of a solid core on the
equatorial flow in giant planets.
{\it Icarus}, 2006, {\bf 181}, 458-464.

\refpar
Evonuk, M. and Glatzmaier, G.A.,
The effects of small solid cores on deep convection in giant planets.
{\it Planet. Space Sci.}, 2007, {\bf 55}, 407-412.

\refpar
Gilman, P.A. and Glatzmaier, G.A.,
Compressible convection in a rotating spherical shell I. Anelastic equations.
{\it Astrophys. J. Suppl.}, 1981, {\bf 45}, 335-349.

\refpar
Glatzmaier, G.A.,
Numerical simulations of stellar convective dynamos. I. The model and method.
{\it J. Comp. Phys.}, 1984, {\bf 55}, 461-484.

\refpar
Glatzmaier, G.A.,
Planetary and stellar dynamos: challenges for next generation models.
In {\it Fluid Dynamics and Dynamos in Astrophysics and Geophysics},
edited by A.M. Soward, C.A. Jones, D.W. Hughes and N.O. Weiss,
Chp. 11, pp. 331-357, 2005 (CRC Press: London).

\refpar
Glatzmaier, G.A.,
A note on ``Constraints on deep-seated zonal winds inside Jupiter and Saturn".
{\it Icarus}, 2008, in press.

\refpar
Glatzmaier, G.A. and Gilman, P.A.,
Compressible convection in a rotating spherical shell.
III. Analytic model for compressible vorticity waves.
{\it Astrophys. J. Suppl.}, 1981, {\bf 45}, 381-388.

\refpar
Glatzmaier, G.A. and Gilman, P.A.,
Compressible convection in a rotating spherical shell.
V. Induced differential rotation and meridional circulation.
{\it Astrophys. J.}, 1982, {\bf 256}, 316-330.

\refpar
Guillot, T.,
A comparison of the interiors of Jupiter and Saturn.
{\it Planet. Space Sci.}, 1999, {\bf 47}, 1183-1200.

\refpar
Guillot, T.,
The interiors of giant planets: Models and outstanding questions.
{\it Ann. Rev. Earth Planet. Sci.}, 2005, {\bf 33}, 493-530.

\refpar
Hammel, H.B., de Pater, I., Gibbard, S., Lockwood, G.W. and Rages, K.,
Uranus in 2003: Zonal winds, banded structure, and discrete features.
{\it Icarus}, 2005, {\bf 175}, 534-545.

\refpar
Hart, J.E., Glatzmaier, G.A. and Toomre, J.,
Spacelaboratory and numerical simulations of thermal convection
in a rotating hemispherical shell with radial gravity.
{\it J. Fluid Mech.}, 1986, {\bf 173}, 519-544.

\refpar
Heimpel, M., Aurnou, J. and Wicht, J.,
Simulation of equatorial and high-latitude jets on Jupiter
in a deep convection model.
{\it Nature}, 2005, {\bf 438}, 193-196.

\refpar
Hubbard,, W.B., Nellis, W.J., Mitchell, A.C., Holmes, N.C.,
Limaye, S.S. and McCandless, P.C.,
Interior structure of Neptune: Comparison with Uranus.
{\it Science}, 1991, {\bf 253}, 648-651.

\refpar
Hubbard,, W.B., Guillot, T., Marley, M.S., Burrows, A.,
Lunine, J.I., Saumon, D.S.,
Comparative evolution of Jupiter and Saturn.
{\it Planet. Space Sci.}, 1999, {\bf 47}, 1175-1182.

\refpar
Ingersoll, A.P. and Pollard, D.,
Motion in the interiors and atmospheres of Jupiter and Saturn:
Scale analysis, anelastic equation, barotropic stability criterion.
{\it Icarus}, 1982, {\bf 52}, 62-80.

\refpar
Ingersoll, A.P., Gierasch, P.J., Banfield, D., Vasavada, A.R.
and the Galileo Imaging Team,
Moist convection as an energy source for the large-scale motions
in Jupiter's atmosphere.
{\it Science}, 2000, {\bf 403}, 630-632.

\refpar
Liu, J., Goldreich, P.M. and Stevenson, D.J.,
Constraints on deep-seated zonal winds inside Jupiter and Saturn.
{\it Icarus}, 2008, in press.

\refpar
Nellis, W.J.,
Metallization of fluid hydrogen at 140 GPa (1.4 Mbar):
implications for Jupiter.
{\it Planet. Space Sci.}, 2000, {\bf 48}, 671-677.

\refpar
Porco, C.C., West, R.A., McEwen, A. {\it et al.},
Cassini imaging of Jupiter's atmosphere, satellites, and rings.
{\it Science}, 2003, {\bf 299}, 1541-1547.

\refpar
Proudman, J.,
On the motion of solids in a liquid possessing vorticity.
{\it Proc. R. Soc. Lond. A}, 1916, {\bf 92}, 408-424.

\refpar
Rhines, P.B.,
Waves and turbulence on a beta-plane.
{\it J. Fluid Mech.}, 1975, {\bf 69}, 417-443.

\refpar
Roberts, P.H.,
On the thermal instability of a rotating-fluid sphere
containing heat sources.
{\it Philos. Trans. R. Soc. London}, 1968, {\bf 263}, 93-117.

\refpar
Rogers, T.M. and Glatzmaier, G.A.,
Penetrative convection within the anelastic approximation.
{\it Astrophys. J.}, 2005, {\bf 620}, 432-441.

\refpar
Rogers, T.M., Glatzmaier, G.A. and Jones, C.A.,
Numerical simulations of penetration and overshoot in the sun.
{\it Astrophys. J.}, 2006, {\bf 653}, 766-773.

\refpar
Sanchez-Lavega, A., Rojas, J.F. and Sada, P.V.,
Saturn's zonal winds at cloud level.
{\it Icarus}, 2000, {\bf 147}, 405-420.

\refpar
Stanley, S. and Bloxham, J.,
Convective-region geometry as the cause of Uranus' and Neptune's
unusual magnetic fields.
{\it Nature}, 2004, {\bf 428}, 151-153.

\refpar
Starchenko, S.V. and Jones, C.A.,
Typical velocities and magnetic field strengths in planetary interiors.
{\it Icarus}, 2002, {\bf 157}, 426-435.

\refpar
Stevenson, D.J.,
Interiors of the giant planets.
{\it Ann. Rev. Earth Planet. Sci.}, 1982, {\bf 10}, 257-295.

\refpar
Sun, Z.-P., Schubert, G. and Glatzmaier, G.A.,
Banded surface flow maintained by convection in a model of the
rapidly rotating giant planets.
{\it Science}, 1993, {\bf 260}, 661-664.

\refpar
Williams, G.P.,
Planetary circulations: 1. Barotropic representation of Jovian
and terrestrial turbulence.
{\it J. Atmos. Sci.}, 1978, {\bf 35}, 1399-1424.

\refpar
Williams, G.P.,
Jovian dynamics. Part III: Multiple, migrating, and equatorial jets.
{\it J. Atmos. Sci.}, 2003, {\bf 60}, 1270-1296.

\appendices
\section{Numerical method}

Our models of rotating thermal convection are based on the anelastic equations
of motion (e.g. Gilman and Glatzmaier 1981, Braginsky and Roberts 1995),
which describe subsonic flow relative to a frame of reference rotating at
angular velocity, ${\bf \Omega}$, and small thermodynamic perturbations relative to
a depth-dependent, hydrostatic, adiabatic reference state
(here denoted by an over-bar).  The equations governing conservation of
mass (A.1) and momentum (A.2) and the equation for heat transfer (A.3) are

\begin{equation}
\nabla \cdot (\bar{\rho}{\bf v}) = 0
\end{equation}

\begin{equation}
\frac{\partial{\bf v}}{\partial t} =
- ({\bf v \cdot \nabla}){\bf v}
- \nabla (\frac{p}{\bar{\rho}}+U)
+ \bar{{\bf g}} \left( \frac{\partial \bar \rho}{\partial S} \right)_p  S
+ 2 {\bf v} \times {\bf \Omega}
+ \nu \nabla^2{\bf v}
\end{equation}

\begin{equation}
\frac{\partial S}{\partial t} =
- ({\bf v \cdot \nabla}) S
+ \frac{\kappa}{\bar{\rho} \bar{T}}
{\bf \nabla \cdot} (\bar{\rho} \bar{T} {\bf \nabla} S) ~~,
\end{equation}

\noindent where ${\bf v}$ is fluid velocity, $\bar \rho$ is reference state density and
$S$, $p$ and $U$ are the perturbations in specific entropy, pressure and
gravitational potential, respectively.  We are using the
co-density formulation, which combines the pressure gradient,
the pressure contribution to the density perturbation and
the gravitational potential perturbation into the gradient of
a reduced-pressure, $(p/\bar\rho)+U$ (Braginsky and Roberts 1995).
To keep this model relatively simple, we have
prescribed constant viscous, $\nu$, and thermal, $\kappa$, diffusivities, which
represent subgrid-scale turbulent diffusion; this is the reason our diffusive heat flux
is proportional to the entropy gradient.  No hyperdiffusion is employed; that is,
these diffusion coefficients do not depend on length scales or wavenumbers.
In addition, to keep this model simple
we have neglected the additional viscous terms in equation (A.2) that
depend on the radial derivative of the
reference state density; and in (A.3) we have neglected viscous heating.
This should not noticeably affect the results of these low viscosity simulations.

Our 2D model is written in cylindrical coordinates ($r, \phi , z$).  It
assumes no flows or gradients in the $z$ direction, the direction of the
rotation rate, ${\bf \Omega} = \Omega {\bf \mbox{\^{z}}}$.
The reference state density, $\bar \rho$, is prescribed as simple functions of $r$.
For Case 1 $\bar \rho = c_1-c_2 r^2$ and for Case 2 $\bar \rho = c_3 r^{-1.25}$,
where $c_1$, $c_2$ and $c_3$ are constants.
Consequently, the density scale height,
$L_{compress}=-(d ~ \mathrm{ln} (\bar \rho) / dr)^{-1}$,
decreases with radius for Case 1 and increases with radius for Case 2.
Integrating density in radius provides the radially-dependent
reference state gravitational acceleration,
$\bar{{\bf g}} = - \bar g {\bf \mbox{\^{r}}}$ in (A.2).
The hydrostatic equilibrium equation
can then be integrated to get the reference state pressure.  An equation of
state provides the reference state temperature, $\bar T$, and
the thermodynamic derivative
$( \partial \bar \rho / \partial S )_p$.  Here we assume a
perfect gas, which makes this thermodynamic derivative equal to
$-\bar{\rho}/c_p$, where $c_p$ is the (constant)
specific heat capacity at constant pressure.

Since this is a 2D problem, we choose to solve the vorticity equation, i.e.,
the curl of (A.2).  This removes the reduced pressure perturbation term
from the calculation.  The resulting equation for the z-component of vorticity,
$\omega = \left( {\bf \nabla} \times {\bf v} \right)_z$, is

\begin{equation}
\frac{d \omega}{dt} =
\frac{\partial \omega}{\partial t} + ({\bf v \cdot \nabla}) \omega =
\frac{\bar{g}}{r \bar{\rho}} \left( \frac{\partial \bar \rho}{\partial S} \right)_p 
\frac{\partial S}{\partial \phi}
- (2 \Omega + \omega) {\bf \nabla \cdot v}
+ \nu \nabla^2 \omega ~~.
\end{equation}

\noindent The compressional torque in this equation is, using equation (A.1),

\begin{displaymath}
- (2 \Omega + \omega) {\bf \nabla \cdot v} = (2 \Omega + \omega) h_\rho v_r ~~,
\end{displaymath}

\noindent where $h_\rho = d ~ \mathrm{ln} (\bar \rho) / dr$, the negative of
the inverse density scale height, $L_{compress}$.
The stretching torque,

\begin{displaymath}
(2 {\bf \Omega} + {\bf \omega}) {\bf \cdot \nabla} v_z ~~,
\end{displaymath}

\noindent vanishes in this 2D model.
We then define a streamfunction, $\Psi$, such that

\begin{equation}
\bar \rho {\bf v} = {\bf \nabla} \times (\Psi {\bf \mbox{\^{z}}})~~,
\end{equation}

\noindent which ensures mass conservation (A.1).  Vorticity is therefore

\begin{equation}
\omega = 
-\frac{1}{\bar \rho} \left[ \frac{\partial^2 \Psi}{\partial r^2}
+ \left( \frac{1}{r} - h_\rho \right) \frac{\partial \Psi}{\partial r}
+ \frac{1}{r^2} \frac{\partial^2 \Psi}{\partial \phi^2} \right]~~.
\end{equation}

We make the inner and outer boundaries impermeable and stress-free.
The former requires $v_r$ to vanish and, by (A.5), $\Psi$ to be constant
on these boundaries.  We therefore set $\Psi = 0$ on both boundaries to make the
total integrated momentum vanish relative to the rotating frame.
The latter boundary condition
requires the radial gradient of $(v_r / r)$ to vanish and therefore

\begin{displaymath}
\left[ \frac{\partial^2 \Psi}{\partial r^2}
- \left( \frac{1}{r} + h_\rho \right) \frac{\partial \Psi}{\partial r} \right] =0
\end{displaymath}

\noindent on both boundaries.  We also constrain entropy to be constant on
the boundaries; the drop in entropy between the boundaries, $\Delta S$,
drives the convection.

We expand $S$, $\omega$ and $\Psi$ in Fourier functions
of $\phi$ and Chebyshev polynomials of $r$.
Equations (A.3), (A.4) and (A.6) are solved simultaneously,
one complex Fourier mode at a time, using Chebyshev collocation.
The time integration treats all linear terms implicitly (including
the Coriolis term).  The nonlinear advection terms are treated explicitly
with the Adams-Bashforth scheme.
A spectral transform method, using fast Fourier transforms in both longitude
and radius, is employed to calculate the nonlinear terms at each time step.
This method is similar to that described in Glatzmaier (1984).

For these anelastic equations,
the numerical time step is limited by the CFL condition based on the fluid
velocity and the spatial grid.  The Chebyshev grid in radius, being much finer
near the boundaries, provides enhanced resolution of the shallow boundary
layers without a severe CFL constraint since the radial velocity vanishes at
the boundaries.

The code is parallel.  For part of each numerical time step the Fourier modes are
distributed over the processors so each of the Chebyshev collocation matrix solutions
can be done within one processor.  During the other part of the step the radial
levels are distributed so each of the spectral transforms in longitude
can be done within one processor.
This requires global transposes each numerical time step, a price one pays
for the enhanced accuracy of spectral methods over local finite difference
methods.

\section{Potential vorticity theorem}

\noindent This theorem is based on the assumption that buoyancy and
viscous terms can be neglected to first order in the vorticity equation (A.4).
However, unlike the Proudman-Taylor Theorem, the inertial term,

\begin{displaymath}
\frac{d \omega}{dt} = \frac{\partial \omega}{\partial t}
+ ({\bf v \cdot \nabla}){\omega} ~~,
\end{displaymath}

\noindent is retained.
Since $\bar \rho$ depends only on $r$, $h_\rho v_r = d ~ ln \bar \rho / dt$.
Also, since $\Omega$ is constant in space and time,
$\partial \omega / \partial t =
\partial (2 \Omega + \omega) / \partial t$.
Substituting these into (A.4) gives

\begin{displaymath}
\frac{d}{dt} \left( \frac{2 \Omega + \omega}{\bar \rho} \right) = 0~~.
\end{displaymath}

\noindent That is, the potential vorticity of a fluid parcel, 
$(2 \Omega + \omega) / \bar \rho$, remains constant
as the parcel moves.
In 3D one would need to neglect the local vortex stretching term
to arrive at this result.  This, however, may not be negligible in a turbulent
density-stratified environment.

\end{document}